\RequirePackage[utf8]{inputenc}
\documentclass[conference]{IEEEtran}

\ifCLASSINFOpdf
  \usepackage[pdftex]{graphicx}
\else
\fi

\usepackage[T1]{fontenc}
\usepackage[numbers,sort&compress]{natbib}
\usepackage[hyphens]{url}
\usepackage{hyperref}
\usepackage{cleveref}
\usepackage{booktabs}
\usepackage{microtype}

\newcommand{\sectionref}[1]{$\S$\ref{#1}}

\def\ptitle{K-resolver: Towards Decentralizing\\ Encrypted DNS Resolution}
\def\pkeywords{}

\hypersetup{
	colorlinks	= true,
	citecolor	= black,
	filecolor	= black,
	linkcolor	= black,
	urlcolor	= black,
	pagebackref	= false,
	pdftitle	= {\ptitle},
	pdfkeywords	= {\pkeywords}
}

\begin{document}
\title{\ptitle}



%

\author{
  \IEEEauthorblockN{Nguyen Phong Hoang, Ivan Lin, Seyedhamed Ghavamnia, Michalis Polychronakis}
  \IEEEauthorblockA{Stony Brook University, New York, USA}
  \IEEEauthorblockA{\{nghoang, ivlin, sghavamnia, mikepo\}@cs.stonybrook.edu}
}


\IEEEoverridecommandlockouts
\makeatletter\def\@IEEEpubidpullup{6.5\baselineskip}\makeatother
\IEEEpubid{\parbox{\columnwidth}{
    {\fontsize{7.5}{7.5}\selectfont Workshop on Measurements, Attacks, and Defenses for the Web (MADWeb) 2020 \\
    23 February 2020, San Diego, CA, USA \\
    ISBN 1-891562-63-0 \\
    https://dx.doi.org/10.14722/madweb.2020.23009 \\
    www.ndss-symposium.org}
}
\hspace{\columnsep}\makebox[\columnwidth]{}}

\maketitle

\begin{abstract}

Centralized DNS over HTTPS/TLS (DoH/DoT) resolution,
which has started being deployed by major hosting providers
and web browsers, has sparked controversy among Internet activists and privacy
advocates due to several privacy concerns. This design decision causes the
trace of all DNS resolutions to be exposed to a third-party
resolver, different than the one specified by the user's access network.
In this work we propose \emph{K-resolver}, a DNS resolution
mechanism that disperses DNS queries across \emph{multiple} DoH resolvers,
reducing the amount of information about a user's browsing
activity exposed to each individual resolver. As a result, none of the
resolvers can learn a user's entire web browsing history.
We have implemented a prototype of our approach for
Mozilla Firefox, and used it to evaluate the performance of web page
load time compared to the default centralized
DoH approach. While our \emph{K-resolver} mechanism has some effect on DNS
resolution time and web page load time, we show that this is mainly
due to the geographical location of the selected DoH servers. When more
well-provisioned anycast servers are available, our approach
incurs negligible overhead while improving user privacy.

\end{abstract}

\section{Introduction}

Introduced in 1983, the domain name system (DNS) is a hierarchical naming
system made up of name servers distributed across the Internet~\cite{rfc882}.
DNS maps human-memorable domain names (e.g., \textit{example.com}) to their
corresponding IP address, which are then used
to route and deliver resources between connected devices. For that
reason, DNS is involved in almost every online activity.
The original DNS protocol, however, was not designed with security and privacy
in mind. More specifically, DNS packets are not encrypted or authenticated,
resulting in several security vulnerabilities and privacy risks.
Man-on-the-side (MoTS) attackers can tamper with DNS responses to re-direct
users to malicious hosts~\cite{Kaminsky2008, Alexiou2010, holdonDNS,
Szurdi:usenixsecurity14}, while state-level adversaries can passively monitor
DNS traffic for surveillance or inject forged DNS responses to block unwanted
connections for censorship purposes~\cite{Aryan:2013, Hoang2016:AppsPrivacy,
farnan2016cn.poisoning, hoang:2019:measuringI2P, iclab_SP20}.

As a response to these threats,
DNS over HTTPS (DoH)~\cite{rfc8484} and DNS over TLS
(DoT)~\cite{rfc7858} were recently proposed and are under active
development, with many web browsers and network operators already supporting
them. Despite the use of different protocols to encapsulate DNS messages,
these two techniques have the same purpose of preserving the integrity and
confidentiality of DNS resolutions against adversaries between a client and its
recursive resolver (also referred to as recursor).

Although the benefits provided by DoH/DoT against threats ``under the
recursor'' are clear, they come with the cost of trusting the DoH/DoT
operator with the entire web browsing history of users. Recently, the design decision
of Firefox~\cite{firefoxDoH}---one of the most popular web browsers---to centralize
DNS resolutions to a single DoH server operated by Cloudflare, has led to
controversy due to several privacy concerns~\cite{bert_RIPEblog,
bert_powerDNS}. A similar plan by Google Chrome has also faced strong
opposition~\cite{Google_DoH_complaint}.

As a step towards mitigating the privacy concerns stemming from the
exposure of \emph{all} DNS resolutions of a user---effectively the user's
entire domain-level browsing history---to an \emph{additional} third party entity,
in this work we propose \emph{K-resolver}, a resolution mechanism in which
DNS queries are dispersed across multiple (\emph{K}) DoH servers, allowing
each of them to individually learn only a \emph{fraction} (\emph{1/K})
of a user's browsing history.

We evaluated the privacy benefit provided by this mechanism
using 200 user profiles, with each profile containing 100 domain names sampled
from the top one million popular websites~\cite{LePochat2019}.
In addition, we also evaluated the impact of our resolution scheme on the
performance of web page load time. Although both DNS resolution and web page
load time take slightly longer when browsing with our \emph{K-resolver} mechanism, we
show that this impact is due to the geographical location of the selected DoH
recursors. If more well-provisioned anycast DoH/DoT servers are
available, our proposed mechanism can be expected to provide comparable
performance while also improving the online privacy of Internet users.

\section{Background and Motivation}
\label{sec:motivation}

\begin{figure}[t]
	\centering
	\includegraphics[width=\columnwidth]{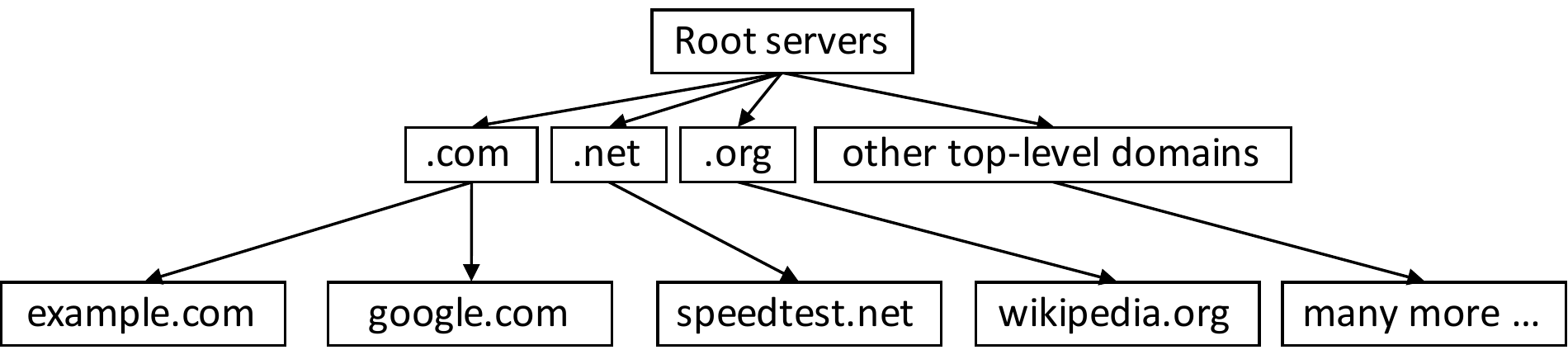}
	\caption{The domain name system hierarchy.}
	\label{fig:dns_hierarchy}
\end{figure}

DNS maps domain names to their IP address(es). The mapping between a domain
and its associated IP address(es) is called a \emph{resource record}, which is
stored in a distributed fashion on the DNS hierarchy, as illustrated in
Figure~\ref{fig:dns_hierarchy}.

There are two traditional resolution methods to obtain the IP address(es) of a
domain. The first way is via an iterative DNS resolution. For instance, to
find the IP address of \emph{example.com}, the client first queries the
\emph{root} name server to obtain the IP address(es) of the \emph{.com} name
server, then queries the \emph{.com} name server for the IP address(es) of
\emph{example.com}, and so on.
However, traversing the whole DNS hierarchy for every
lookup can be costly and time consuming if the client needs to look up many
different domains multiple times. Recursive DNS resolution was introduced to
cope with this problem. In a recursive resolution, a \emph{recursor} acts as a proxy
between the client and the rest of the DNS hierarchy. When the recursor
receives the query for \emph{example.com} from the client, it will
iteratively look up the IP address(es) in the same manner described above if
the resource record of \emph{example.com} is not in its cache. However, if
the domain was queried by another client and was previously cached, the recursor
will answer the query directly from its cache, saving the iterative
lookup overhead.

As mentioned earlier, these traditional DNS resolutions are not encrypted,
exposing all DNS queries and responses in plaintext, and thus leading to
numerous security and privacy problems. The DoH and DoT protocols were
proposed as an effort
to address these problems. While these two techniques can be deployed at all
levels of the DNS hierarchy to encrypt DNS messages between the involved parties,
there is recently a push for the deployment of DoH/DoT ``under the
recursor,'' with major public DNS resolvers already supporting it (e.g.,
Google~\cite{Google_DoT}, Cloudflare~\cite{cloudflare1111}). Although the
deployment of DoH/DoT ``above the recursor'' at authoritative name servers is
also recommended by the Internet Engineering Task Force~\cite{IETF106}, it has
not been widely adopted. To the best of our knowledge, at the time of writing
this paper, only the name server of Facebook has collaborated with Cloudflare
to secure DNS traffic between Cloudflare's recursors and Facebook's name
servers~\cite{cloudflare_FB}.

Encrypting DNS messages to protect them from eavesdropping or manipulation
is a desirable
improvement from which all privacy-conscious Internet users would benefit.
However, the recent controversy is over how that encryption should be done,
especially when it comes to the cost of having to trust an \emph{additional} entity with
the user's entire browsing history,
in case the web browser is configured to centralize all
DNS resolutions to only one recursor---which is currently the default
configuration for Mozilla Firefox~\cite{firefoxDoH}.
This motivated us to explore the \emph{K-resolver} approach
as a resolution mechanism to remedy this ``single-point-of-trust'' problem.
More specifically, instead of sending all DNS resolution requests to only one
recursor, our mechanism allows users to distribute their DNS queries among
multiple \emph{K} recursors. As a result, none of the recursors can learn the
entire browsing history of the user.

\section{Threat Model}
\label{sec:threat}

The bottom line of recent debates around DoH/DoT is about the decision of
centralizing all DNS queries into a single DoH/DoT third-party recursor by
major browser vendors (Mozilla Firefox from version 62~\cite{firefoxDoH}, and
Google Chrome from version 79~\cite{Google_DoT}). While users can configure
these browsers to use any DoH recursor of their preference, they still have to
trust the selected recursor with their entire web browsing history, posing a
severe threat to their privacy if the recursor maliciously uses the observed
DNS queries for illicit purposes.

To mitigate this privacy threat, our \emph{K-resolver} mechanism is proposed
to limit the amount of browsing information known by each recursor. Note that
our proposal does not aim to anonymize a given user's browsing activities by
decoupling the link between users and their DNS queries, which is not the
original purpose of DoH/DoT either. For anonymous DNS resolutions, users may
use anonymous communication systems to route their DNS
traffic~\cite{Hoang2014:Anonymous}. For instance, DNS-over-HTTPS-over-Tor is
one of the more complicated methods to secure DNS traffic, which has been
already implemented by Cloudflare~\cite{cloudflareDNS-Tor}.

Our threat model is oblivious to the relationship between hosting providers
and operators of the recursor, who may be the same entity. Nonetheless, we
consider this relationship as an orthogonal problem due to the current
state of web co-location, in which the vast majority of web servers are
hosted by only a handful of hosting providers~\cite{Shue:2007}, among which
Google and Cloudflare are dominant~\cite{hoang2019_dns_privacy}. For example,
with our \emph{K-resolver} mechanism, the domain name of a website may be
resolved by a recursor that does not belong to Google or Cloudflare. However,
if the website or a third-party resource of the site happens to be hosted by
Google or Cloudflare, then the fact that it was visited by a particular user
cannot be hidden from these same companies.

Ultimately, with the \emph{K-resolver} resolution mechanism in place, each
among \emph{K} selected recursors can only obtain at most \emph{1/K} part of
the user's browsing history, which is still a desirable outomce for
privacy-conscious users.

\section{Design}
\label{sec:design}

As most privacy concerns around DoH stem from the design decision
of major web browsers to
centralize all DNS queries into a single DoH recursor, we focused on
implementing our
proposed mechanism in a web browser as a proof of concept.
Since Mozilla is one of the first vendors to support DoH in their browser, we
chose to use Firefox and modify its source code to integrate our
\emph{K-resolver} mechanism into the browser. To that end, our implementation
of the \emph{K-resolver} mechanism is solely a client-end modification of the
Firefox web browser without changing any other components of the DNS hierarchy
or the DoH/DoT standards.

The design of our
\emph{K-resolver} approach is centered around two main design considerations,
which we discuss below.

\subsection{Domain-recursor Pair Fixation}

A particular domain name, which is
entered in the URL bar of the browser (parent domain), should always be resolved
by the same recursor. One of the ways to distribute DNS queries is naively
rotating queries among recursors in a round-robin fashion. If the number of
visited websites is small while the number of recursors is large enough,
a round-robin rotation approach may achieve the design goal of our
\emph{K-resolver} mechanism over a short period of time.
Unfortunately, round-robin
rotation is not a sustainable solution as all selected recursors will eventually observe
all domain names visited over a long period of time, making the situation even
worse. Therefore, a particular parent domain should always be resolved by the
same recursor, preventing other recursors from learning the visit to that
domain.

Given an indexed list of \emph{K} DoH recursors selected by the user,
a domain name is resolved by a recursor whose index is calculated by taking
the modulo of the hashed value of that domain name concatenated with a salt
value, with respect to the \emph{K} number of selected recursors. For each
user, the salt value, which can be configured by and is only known by that user, is
generated randomly so that two different users using the same list of
recursors will not end up with selecting the same recursor to resolve the same
domain.

\subsection{Parent-domain-based Recursor Selection}

All DNS queries generated
from within the same web page should be resolved by the same recursor of the
parent domain. Rendering a web page often requires the download of several third-party
resources, including images, scripts, and style sheets, as observed by previous
work~\cite{Nikiforakis2012, falahrastegar2014, Lauinger2017, Mueller2019,
Ikram2019, Patil:ANRW2019, Bottger:IMC2019}. DNS resolutions of these
third-party resources can be used as an effective feature to fingerprint web
sites. Even when being split across recursors, some unique third-party domain
names can potentially be used to trace back to the parent domain. Therefore,
all the DNS queries generated due to the resolution of third-party elements of
a given page should also be
resolved by the same recursor of the parent domain, instead of sending them to
different resolvers.

\section{Experiment Setup}
\label{sec:experiment}

In this section, we  discuss how we set up and conduct experiments to
analyze the privacy benefit of our \emph{K-resolver} mechanism and its impact
on web page load time.

\subsection{Domain Name Dataset}

We first generate 200 user profiles with each of them containing 100 unique
domain names. We approximate this number based on the result from a prior
study showing that most Internet users visit an average of 89 domains per
month~\cite{domain_per_month}. We sample the domains from the top list of one
million popular websites obtained from the Tranco project on December 12th,
2019~\cite{LePochat2019}. To simulate a real-world scenario, for each user
profile, we select the domains based on their popularity ranking instead of
randomly picking them from the top list. Hence, a domain with a higher rank is
more likely to be selected.

\subsection{Selection of DoH Recursors}

Next, we use public DoH servers provided by the Curl project to curate a list
of recursors~\cite{curl_DoH}. At the time of conducting our study, there are
38 operators offering 53 recursors. However, many of these recursors integrate
filtering capabilities, including parental control and advertisement blocking.
Moreover, some recursors are located in countries that are notorious for
Internet censorship---for instance, \textit{Rubyfish.cn}'s recursor is located
in China. To make sure that DNS resolutions are not biased and tampered with
throughout our experiment, we only select those active recursors that do not
filter or censor any content. After excluding recursors that are not suitable
for our study, we were left with a curated list of 26 DoH recursors.

\subsection{Automated Web Crawling and Performance Measurement}

Finally, with the
\emph{K-resolver} mechanism integrated in the Firefox browser, we visited 100
domain names with each user profile generated above. For every domain name, we
launch a new Firefox session to clean cached contents of previously visited
domains so that they will not have any impact on the performance metrics of
the domain currently being visited.
The whole process, from when the browser is launched
until the web page is loaded, is handled by the Browsertime
framework~\cite{Browsertime}.

After completely loading the web page,
Browsertime is configured to export an HTTP Archive (HAR file), containing
several performance metrics which we later use for our analyses
in~\sectionref{sec:pageloadtime}. Furthermore, to account for unavoidable
factors that may affect our experiments, such as downgraded network performance
of up-stream providers or web servers of tested domains being busy, we visited
each domain at three different times, from the 18th to the 24th of December,
2019. We conducted our experiments on machines running Ubuntu 18.04.3 LTS with
32GB of RAM, 8 Intel Xeon X5450 processors, connected to a Gigabit educational network
located in the East coast of the United States.

\section{Data Analysis}
\label{sec:analysis}

Using the data collected in~\sectionref{sec:experiment}, we analyze the
privacy benefits provided by our \emph{K-resolver} mechanism and its impact on
web page load time, compared to the default setting of Cloudflare's DoH
recursor in the Firefox browser. For each of the 26 DoH recursors selected, we
also investigate the effect that their geographical location has on DNS
resolution time.

\subsection{Privacy Benefit Analysis}

\begin{figure}[t]
\centering
\includegraphics[width=0.9\columnwidth]{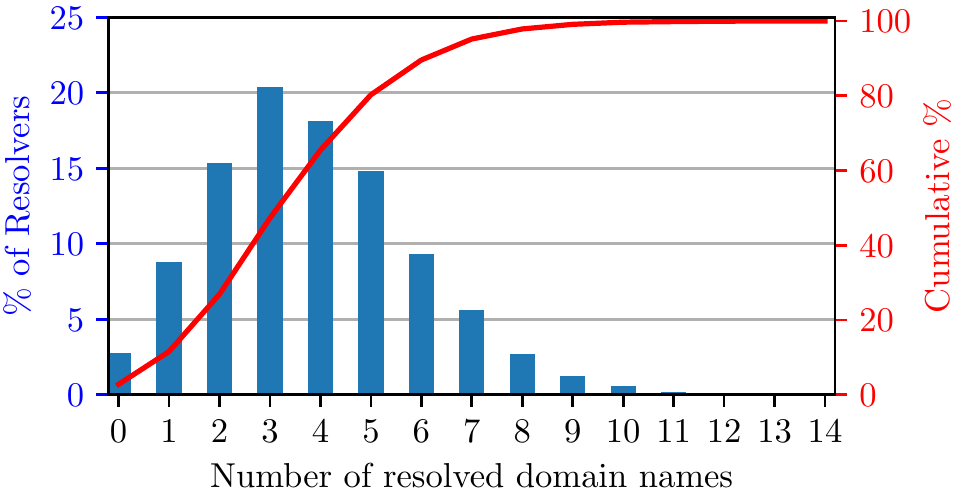}
\caption{CDF of domains resolved per DoH recursor.}
\label{fig:resolved_domains_per_recursor}
\end{figure}

As introduced in~\sectionref{sec:threat}, the privacy benefit of our
\emph{K-resolver} mechanism is obtained by uniformly distributing one's
visited domains among \emph{K} recursors, such that each recursor only learns a
\emph{1/K}-portion of the user's browsing history. For instance, given 100
domain names per user profile and 26 DoH recursors, each recursor should
ideally resolve just 3--4 domain names.

However, in practice it is extremely hard to obtain an even distribution due
to several uncertainties. For example, in a real-world scenario, we cannot
know all domains that a user will visit in advance, and thus cannot guarantee
that our bucket-hashing algorithm will evenly split all domains across the
selected recursors. Figure~\ref{fig:resolved_domains_per_recursor} shows the
distribution of the number of domains resolved per recursor in our experiment.
The number of domains resolved by each recursor exhibits a skewed normal
distribution, in which most DoH recursors (40\%) resolve 3--4 domains.
However, there are some corner cases in which 2.8\% of recursors do not
resolve any domains, while 2.2\% of the recursors resolve from 8 to 14
domains.

To examine whether our bucket-hashing algorithm can evenly disperse
\emph{sensitive} domains across recursors, we consider some domain categories as
``sensitive'' and check if the set of sensitive domains within the same user profile are
resolved by different DoH recursors (the desirable behavior).
More specifically, we use Fortiguard's
Web Filter Lookup~\cite{fortinet} to categorize the domain names selected
in~\sectionref{sec:experiment}, then consider the following categories as
sensitive: pornography, armed forces, gambling, health and wellness, illegal
or unethical, malicious websites, medicine, other adult materials,
plagiarism, proxy avoidance, spam URLs, and weapons (sales).

Among the 200 user profiles generated in~\sectionref{sec:experiment}, we have 182
profiles containing from 1 to 6 sensitive domains each. Among these profiles,
there are only 23 profiles having sensitive domains resolved by different
recursors, while 159 profiles have more than one sensitive domain resolved by
the same recursor.

One may consider that having more than one sensitive domain resolved by one
DoH recursor is harmful for privacy. However, this is a result of the uncertainty
regarding the sensitivity of a given website visited by a user,
of which we do not have any control.
Moreover, the sensitivity of a website can vary from site to site, depending
on who, when, and from where is visiting the
site~\cite{Hoang2016:AppsPrivacy}. Therefore, our current design of the
\emph{K-resolver} mechanism is oblivious to the sensitivity of the resolved
domains. Future work may take this sensitivity aspect into consideration when
assigning (hashing) the domain name to a given recursor.

\subsection{Web Page Load Time Evaluation}
\label{sec:pageloadtime}

Next, we assess the impact of the \emph{K-resolver} mechanism on~i) the DNS
resolution time, and~ii) the web page load time.

\begin{figure}[t]
\centering
\includegraphics[width=0.9\columnwidth]{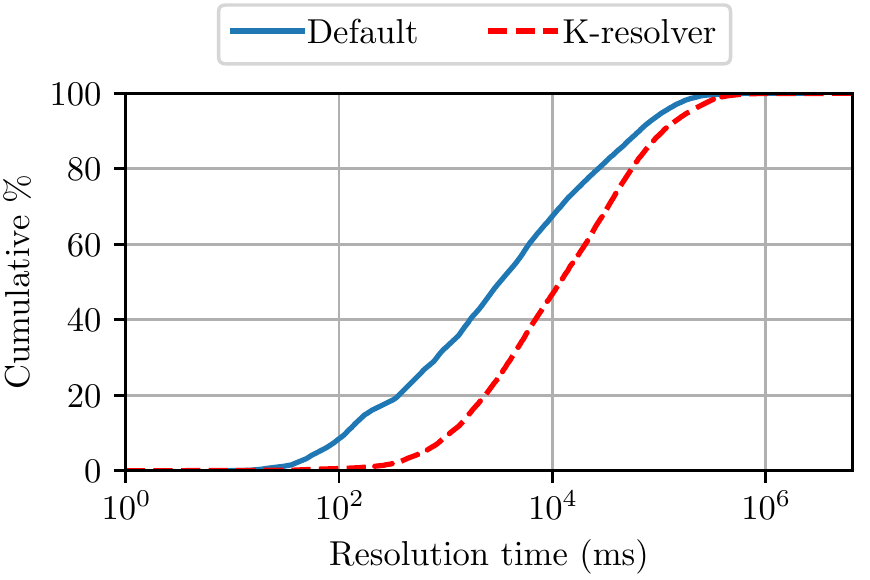}
\caption{CDF of the total DNS resolution time.}
\label{fig:sum_DNS_cdf}
\end{figure}

Figure~\ref{fig:sum_DNS_cdf} shows the CDF of the total amount of time (in
milliseconds) that it takes for all DNS names per website visit to be resolved. The
default (continuous) line represents the time when browsing with the default
DoH setting of Firefox, while the \emph{K-resolver} (dashed) line represents the
time when browsing with our \emph{K-resolver} mechanism enabled.

B\"{o}ttger et al. conducted a measurement study in
April 2019 to compare the resolution time between the traditional DNS and DoH
for the 1,000 most popular Alexa sites~\cite{Bottger:IMC2019}. Their result shows
that more than 50\% of web page visits take longer than 5,000ms in total to
resolve all domains with Cloudflare's DoH recursor. Our experiment with the
default DoH setting of Firefox, however, shows that the total resolution time
has been reduced. Specifically, more than 50\% of our web page visits require
less than 3,600ms in total to resolve all domains.

Our \emph{K-resolver} mechanism has a longer total resolution time
compared to the original DoH setting.
More than 50\% of all visits have a total DNS resolution time longer than
10s, about three times more than the default DoH setting of Firefox. This is an
expected result as not all DoH recursors we selected
in~\sectionref{sec:experiment} support anycast, while many of them are located in
a different continent than our vantage point (USA). We
further investigate the impact of the geographical location of these DoH
recursors on the resolution time in~\sectionref{sec:recursor_location}.

When rendering a website, the DNS queries of involved domains can be issued in a
parallel fashion~\cite{Google_prefetching, firefox_DNS_prefetching},
depending on various uncontrollable factors (e.g., the underlying operating
system, network conditions). Moreover, websites often have a
different number of resources hosted on numerous domains. Therefore, in many
cases, focusing only
on the total amount of time for all DNS queries to be resolved
will not reflect the actual resolution time in real-world
scenarios. Since B\"{o}ttger et al.~\cite{Bottger:IMC2019} only report the
total resolution time per website, we also capture the response time of each
individual DNS resolution under both settings, i.e., the default DoH setting
and the \emph{K-resolver} mechanism.

\begin{figure}[t]
\centering
\includegraphics[width=0.9\columnwidth]{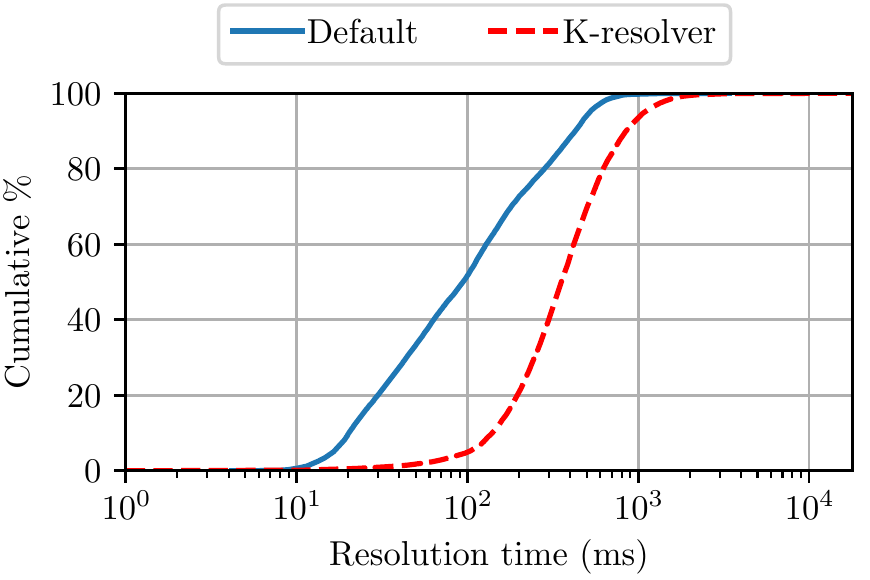}
\caption{CDF of the average DNS resolution time.}
\label{fig:avg_DNS_cdf}
\end{figure}

The CDF in Figure~\ref{fig:avg_DNS_cdf} shows the average time that each
individual DNS resolution takes when resolving with the default DoH setting
and with our \emph{K-resolver} mechanism. When using Cloudflare's DoH recursor
in the default Firefox setting, we observe that 50\% of DNS resolutions take
less than 100ms, while almost all DNS queries are resolved within 1,000ms. In
contrast, when resolving with our \emph{K-resolver} mechanism, less than 5\%
of the resolutions finish within 100ms, while more than 10\% take
longer than 1,000ms to be resolved. We further explain this discrepancy
in~\sectionref{sec:recursor_location}.

DNS resolution time directly impacts the web page load time, as domains
embedded in the web page need to be resolved before the web browser can fetch
resources hosted on these domains. Next, we analyze the web page load time of
domains in the 200 user profiles generated in~\sectionref{sec:experiment}. The
web page load time is defined as the time until the ``onLoad'' event is
triggered.

\begin{figure}[t]
\centering
\includegraphics[width=\columnwidth]{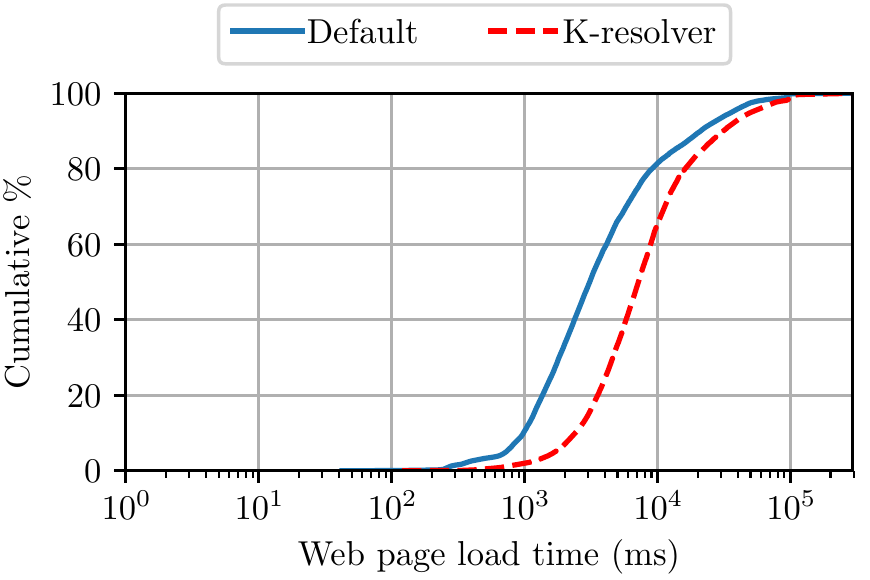}
\caption{CDF of the web page load time.}
\label{fig:onLoad_cdf}
\end{figure}

Due to the longer DNS resolution time as shown in
Figures~\ref{fig:sum_DNS_cdf} and~\ref{fig:avg_DNS_cdf}, it is expected that
most websites would take a longer time to be loaded when browsing with our
\emph{K-resolver} mechanism. As shown in Figure~\ref{fig:onLoad_cdf}, 80\% of
the visits finish loading the web page within 10,000ms under the default DoH
setting, while only 60\% of the visits are completely loaded within the same
amount of time for our approach.

\subsection{Impact of DoH Recursors' Geographical Location}
\label{sec:recursor_location}

\begin{figure}[t]
\centering
\includegraphics[width=0.9\columnwidth]{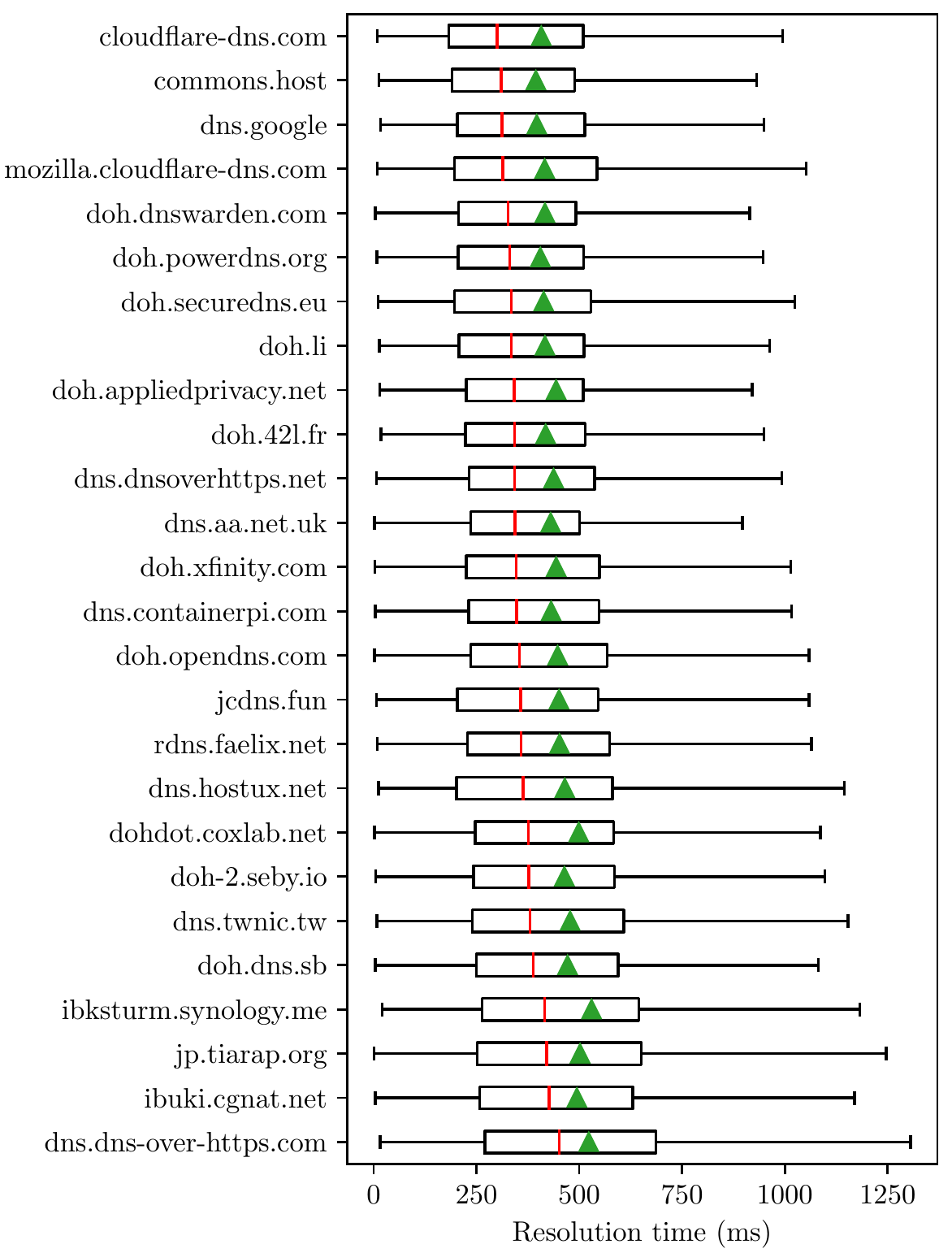}
\caption{The distribution of DNS resolution time of the 26 DoH recursors.}
\label{fig:DoH_recursor_resolution_time}
\end{figure}

Next, we analyze each individual DoH recursor to examine the impact of its
geographical location on the resolution time.
The box plot (excluding outliers) in
Figure~\ref{fig:DoH_recursor_resolution_time} shows the distribution of the
resolution time of the 26 DoH recursors selected
in~\sectionref{sec:experiment}. Both Cloudflare's DoH servers are within the
top five servers with the best performance,
with a median resolution time ranging from 300ms to
330ms. The rest of the recursors have a median resolution time ranging from
330ms to 450ms. This discrepancy in performance among these recursors is
primarily due to their geographical location. More specifically, a closer
recursor to our experimental location (East Coast, USA)
is more likely to provide a lower response latency.

Table~\ref{table:high_privacy_hosting} shows the geographical location, the
ping time, and the median DNS resolution time of the 26 recursors used in our
experiments. Except for those recursors that do not respond to our ping
(ICMP) packets (whose ping time is marked as \textbf{N/A}), there is a
correlation between the geographical location and the ping time. Notably,
most recursors that are hosted on anycast IP address(es) have the shortest
ping time, while unicast recursors have a longer ping time.

\begin{table}[t]
\centering
\caption{Geographical location information of the 26 DoH recursors and the
median of their resolution time.}
\scalebox{1.13}{
\begin{tabular}{llrr}
\toprule
DoH Server & Location   & Ping (ms) & Resolution (ms)\\
\midrule
cloudflare-dns.com         & anycast &   3.14 & 300 \\
commons.host               & anycast &   3.99 & 310 \\ 
dns.google                 & anycast &   2.93 & 312 \\
mozilla.cloudflare-dns.com & anycast &   3.19 & 314 \\
doh.dnswarden.com          & DE      &  87.50 & 327 \\
doh.powerdns.org           & NL      &  77.90 & 331 \\
doh.securedns.eu           & NL      &  85.30 & 335 \\
doh.li                     & UK      &  74.90 & 335 \\
doh.appliedprivacy.net     & AT      &  96.60 & 342 \\
doh.42l.fr                 & FR      &  76.20 & 343 \\
dns.dnsoverhttps.net       & US      &  73.00 & 343 \\
dns.aa.net.uk              & UK      &  71.40 & 344 \\
doh.xfinity.com            & US      &  23.53 & 347 \\
dns.containerpi.com        & JP      & 162.10 & 347 \\
doh.opendns.com            & anycast &   3.90 & 355 \\
jcdns.fun                  & NL      &  85.23 & 358 \\
rdns.faelix.net            & anycast &  82.70 & 358 \\
dns.hostux.net             & LU      &  81.60 & 364 \\
dohdot.coxlab.net          & US      &    N/A & 376 \\
doh-2.seby.io              & AU      &    N/A & 377 \\
dns.twnic.tw               & TW      & 230.00 & 380 \\
doh.dns.sb                 & anycast &   3.20 & 388 \\
ibksturm.synology.me       & CH      & 105.80 & 416 \\
jp.tiarap.org              & anycast &   3.28 & 421 \\
ibuki.cgnat.net            & BR      & 137.10 & 426 \\
dns.dns-over-https.com     & JP      &   N/A  & 452 \\
\bottomrule
\end{tabular}
}
\label{table:high_privacy_hosting}
\end{table}

While the ping time to each recursor can be a good indicator for its response
latency,
the DNS resolution time also depends on how well-provisioned the
recursor is. As shown in Table~\ref{table:high_privacy_hosting}, the top four
anycast recursors with the best resolution time also have relatively small
ping time. However, there are other four anycast recursors also have
relatively small ping time, but are not among the top recursors with shortest
resolution time. This result shows that the impact on the latency of
resolution time and web page load time is mainly because of the geographical
location of the recursors that we select. Assuming an idealistic future
scenario in which more well-provisioned anycast DoH/DoT recursors become
available, our proposed mechanism can be expected to provide comparable
performance to the existing status quo,
while also improving the online privacy of Internet users.

\section{Related Work}
\label{sec:related}

The rapid development and deployment of DoH/DoT has attracted many researchers
to study these new protocols. Recent studies mainly investigate two aspects of
DoH/DoT: privacy~\cite{bushart_padding, Houser:2019:CoNEXT19, pdot,
hoang2019_dns_privacy} and performance~\cite{Lu:2019:IMC19, Bottger:IMC2019,
Deccio:CoNEXT19}.

In terms of user privacy, researchers aim to tackle two types of
adversaries. The first type includes ``nosy'' on-path observers, who have the
privilege to monitor the encrypted traffic between the user and the DoH/DoT
server, trying to predict the website being visited by the user using
different traffic fingerprinting techniques~\cite{bushart_padding,
Houser:2019:CoNEXT19, hoang2019_dns_privacy}. The second type of adversary
is located at the recursor, such as compromised or malicious recursors that have
illicit intentions to use of the user's online history~\cite{pdot}.

Bushart et al.~\cite{bushart_padding} find that,
even with encryption, DNS traffic is still
susceptible to traffic fingerprinting attacks based on packet length and count.
To address this problem, several packet padding
strategies have been implemented~\cite{rfc8467}.
Siby et al.~\cite{encrypteddns} 
developed a model for classifying and predicting encrypted DoH traffic using
packet sizes as a key feature. The authors show that recently implemented
padding strategies are not effective and still cannot cope with traffic
analysis attacks.

Currently, client-side implementations of DoH/DoT still use a single recursor.
This design, however, turns the selected DoH/DoT recursor into a single point
of failure, but also a single point of trust. For instance, several failures
of Cloudflare's recursor have caused large-scale disruptions in web browsing
activities~\cite{cloudflare_failure}. To protect user data from compromised
resolvers, Nakatsuka et al.~\cite{pdot} developed an architecture based on
trusted execution environments and remote attestation to allow users to verify
the resolution software running on the recursor.

Schmitt et al.~\cite{schmitt_oblivious} propose a resolution mechanism to
anonymize users' DNS queries by decoupling the link between users and their
DNS queries. More specifically, the mechanism introduces an entity, called an
Oblivious DNS (ODNS) resolver, residing between the user and the traditional
recursor. This ODNS resolver hides the identity of the user from the
traditional resolver, while not knowing any information about the domain names
being queried.

While it might be expected that DoT/DoH resolutions will take longer than
traditional DNS due to the need to establish a secure connection, Hounsel et.
al~\cite{hounsel_cost_benefit} conducted a measurement study and found that
this is not always the case. Similarly to our findings
in~\ref{sec:recursor_location}, the authors find that the resolution time can
vary from resolver to resolver depending on the implementations of both the
DNS client and the resolver. In addition to specific resolvers, the authors
find that encrypted DNS can perform just as well or even better than
traditional DNS in certain network conditions. Overhead costs of establishing
a TCP connection could be amortized on large loads, and TCP allows for faster
retransmissions on lossy networks~\cite{hounsel_cost_benefit}.

\section{Discussion}

In our current prototype, we use bucket hashing to disperse DNS resolutions
across DoH servers. However, we have shown in~\sectionref{sec:analysis} that
there are still some corner cases that would benefit from future work that
will investigate more case-specific domain resolution distribution algorithms.
More specifically, we observed cases in which several ``sensitive'' domains
are resolved by the same DoH recursor, which is not optimal as long as there
are other resolvers in the pool that have not received any sensitive
resolution at all. However, as mentioned earlier, the sensitivity of a domain
depends on who, when, and from where visited a given website. Therefore,
another domain name distribution approach that future work can investigate is
to take the website category into consideration, such that domains in the same
category are resolved by the same recursors.

As discussed in~\sectionref{sec:design}, all subsequent DNS queries
generated by rendering a web page are also resolved by the same DoH recursor
that resolves the website's parent domain. However, DNS prefetching could
have a negative impact on the privacy benefit provided by our \emph{K-resolver}
mechanism. For example, consider a user who relies on a search engine to look up websites of
interest. If DNS prefetching is enabled, the recursor that resolves the search
engine domain name (e.g., \textit{bing.com}, \textit{google.com}) would also
observe all domain names of interest that the user could potentially visit
later. To that end, we suggest that DNS prefetching should be disabled when
using \emph{K-resolver}, as also is
recommended by prior studies~\cite{Krishnan2010, Krishnan2011}.

While we opt to implement our \emph{K-resolver} mechanism in a web browser
since this is the root cause of recent debates surrounding centralized
DoH/DoT, DNS resolution is generally handled by the operating system (OS).
Although there have been some efforts to support DoH/DoT resolution at the OS
level~\cite{stubby, Microsoft_DoH}, all queries are still sent to a single
``trusted'' resolver. We hope that our proposed mechanism will also prompt OS
vendors to take a major role in adjusting the OS resolver so that multiple
DoH/DoT recursors can be configured at the OS level.

\section{Conclusion}

For more than 35 years, DNS has been one of the core protocols of the
Internet. Its original unencrypted design does not take security and privacy
into consideration, allowing any on-path entity between a user and the DNS
recursor to learn the user's browsing history. The DoH/DoT protocols aim to
solve this very problem by transmitting DNS queries over an encrypted HTTPS/TLS
channel. However, the recent design decision of major web browsers to centralize
all DNS queries into one DoH/DoT recursor has sparked controversy among
Internet activists and online privacy advocates.

In this work, we introduced
\emph{K-resolver}, a resolution mechanism in which DNS queries are dispersed
across multiple recursors, exposing this way only a \emph{fraction} of a user's
browsing history to each recursor. Although the impact of this
resolution mechanism on web page load time is obvious, we have shown that this
is primarily due to the lack of well-provisioned anycast recursors. We hope
that our work will inspire further privacy-enhancing improvements
to the default DoH configuration 
of major browsers, as well as operating systems.
We believe that, in an idealistic future where the number of
independent, well-provisioned anycast recursors will be larger,
distributing DNS queries across multiple
recursors is a promising approach for improving the online privacy of
Internet users.

\section*{Acknowledgments}

We would like to thank the anonymous reviewers for their thorough
feedback on earlier drafts of this paper. We also thank Peter Hedenskog, the
creator of \href{https://sitespeed.io}{sitespeed.io}, for helping us with
debugging the Browsertime framework in a timely manner.

{\footnotesize
\bibliographystyle{IEEEtranS}
\bibliography{main}
}

\end{document}